\def\H{{\cal H}}
\def\h{{1\over 2}}
\def\T{{\mbox{\boldmath$T$}}}
\def\C{{\mbox{\boldmath$C$}}}

\documentstyle[12pt]{article}

\title{\marginpar{\vspace{-1in}\hspace{-1in}\small KFT U{\L} 10/92}
Quantum particle on a quantum circle \thanks{The work
supported by the KBN grant 2 0218 91 01}}
\author{Tomasz Brzezi\'nski \thanks{On leave from University of
{\L}\'od\'z, Institute of Mathematics, ul.\ Banacha 22, 90--238
{\L}\'od\'z, Poland}\\ University of Cambridge
\\ Department of Applied Mathematics\\ and Theoretical Physics \\
CB3 9EW Cambridge, U.K. \and  Jakub Rembieli\'nski
\& Kordian A. Smoli\'nski \\ University of \L\'od\'z \\
Department of Theoretical Physics \\ Pomorska 149/153, 90--236
\L\'od\'z, Poland}
\date{August 1992}
\begin{document}

\maketitle
\begin{abstract} We describe a $q$-deformed dynamical system
corresponding to the quantum free particle moving along the
circle. The algebra of observables is constructed and
discussed. We construct and classify irreducible
representations of the system.
\end{abstract}

\section{Introduction}
Non-commutative geometry \cite{connes2,manin2} has
attracted much attention of theoretical physicists. It is based
on the idea, that the commutative algebra of functions on a
manifold can be replaced by an abstract non-commutative algebra.
This is in the similarity to the usual quantisation procedure,
when one considers a non-commutative algebra of operators in
place of the commutative algebra of real-valued functions as a
set of observables. In particular usual  quantum mechanics can
be understood as a non-commutative symplectic geometry
\cite{dimakis1}.  Another point of view is presented in
\cite{rembielinski1} where quantum dynamics is treated as a
non-commutative diferential calculus (quantum deRham complex).
The ideas of non-commutative geometry lead to the concept of
non-commutative or $q$-deformed physics recently realised as a
number of simple quantum mechanical models with the $q$-deformed
phase-space structure
 \cite{alekseev1,arefeva1,rembielinski1,rembielinski2,wess2}.

 In this paper we construct a simple toy model of the
non-commutative quantum mechanics, i.e.\ we describe the motion
of the quantum  particle on a quantum circle. Our goal is to
describe the unitary time evolution in the non-commutative
phase-space. The similar problem has been stated in
\cite{rembielinski1}, where the motion of the particle on
quantum line has been considered.  There are two possible
schemes of the construction of quantum mechanical
non-commutative models described by the so  called Faddeev's
rectangle. One can begin with the classical system, quantise it
and then deform or one can first perform deformation and then
quantisation. Both procedures do not necessary lead to the same
quantum mechanical system. In  our case however, the Faddeev's
rectangle appears to be commutative.

In the quantum theory one constructs the algebra of observables
$\H(I,x,p)$ which in our case is generated by the hermitean
angular momentum $p$ and unitary position operator $x$,
interpreted as $x = e^{-i\phi}$ where $\phi$  is the angle. They
obey the Heisenberg commutation relation $[x,p]=\hbar x$.
Having defined the algebra of observables one can construct the
notions of states, measurement, mean value etc.\ which are
related to the irreducible Hilbert space representations of the
algebra of observables. The dynamics of the system is given by
the unitary time evolution which is provided by the Heisenberg
equations of motion $\dot{\mit\Omega} = {i \over
\hbar}[H,{\mit\Omega}] + \partial_t{\mit\Omega}$. The possible convenient
description of the dynamics is given by the suitable deRham
complex. The demanding of the unitary time evolution forces the
Heisenberg equations of motion to be unchanged on the $q$-deformed
level too. This implies that the $q$-deformation leaves the
probabilistic interpretation of the system unchanged. The only
part in that scheme which can  be deformed is the algebra of
observables. This is in remarkable contrast to \cite{arefeva1}
where the Heisenberg equations of motion were deformed. In our
case we can demand that the phase-space is given by the quantum
cylinder rather than the classical one.

The paper is organized as follows. In Section 2 we describe the
usual quantum particle on a circle from the algebraic point of
view. In  Section 3 we deform the algebra of observables of the
particle on circle and we solve the Heisenberg equations of
motion showing that the unitary time evolution is possible in
the case of a free particle. Section 4 is devoted to the
construction and classification of the irreducible
representations of the deformed algebra of observables. Finally
in Section 5 we discuss the classical limits and the invariance
of constructed algebra under space and time inversions.

\section{Quantum mechanics of the particle on a circle}

In the standard approach, quantum free particle on a circle is
described by the unitary operator $x$, corresponding to the
position of the particle and an hermitean operator
$p$---canonical momentum (angular momentum). The dynamics of the
system is given by the hermitean Hamiltonian $H = {p^2
\over{2B}}$, where $B$ denotes the moment of inertia of the
particle. The algebra of observables ${\cal H} (I,x,p)$ can be
defined as
\begin{equation}
{\cal H} (I,x,p) = {\C}[I,x,p] / J(I,x,p).
\end{equation}
Here  ${\C}[I,x,p]$ is an associative, involutive (i.e.\ equipped with $^*$
str
ucture, which is represented as the
hermitean conjugation on Hilbert space)  free algebra over $\bf
C$ generated by $I,x,p$ ($I$ is the identity) and $J(I,x,p)$ is
a two-sided ideal in ${\C}[I,x,p]$ generated by the relation:
\begin{equation}
xp = px +\hbar x.
\end{equation}
Notice that the parameters of the theory (e.g.\ $B$ or $\hbar I$)
can be treated as  independent of time operators belonging to
the center of the algebra of observables. Namely we can extend
the algebra $\cal H$ to the algebra ${\cal H}'$ defined by
\begin{equation}
{\cal H}' = {\C}[I,x,p,K,{\mit\Lambda}]/J(I,x,p,K,{\mit\Lambda})
\end{equation}
where the new generators ${\mit\Lambda}, K$ are hermitean and
$J(I,x,p,K,{\mit\Lambda})$ is a two-sided ideal defined by the
relations
\begin{eqnarray}
xp & = & px + \hbar{\mit\Lambda} x \nonumber \\
x{\mit\Lambda} & = & {\mit\Lambda} x \nonumber \\
p{\mit\Lambda} & = & {\mit\Lambda} p \nonumber \\
xK & = & Kx  \label{def.j}\\
pK & = & Kp  \nonumber \\
{\mit\Lambda} K & = & K {\mit\Lambda}. \nonumber
\end{eqnarray}
${\mit\Lambda}$ and $K$ are assumed both invertible and
${\mit\Lambda}$ is positive definite. The Hamiltonian reads
\begin{equation}
H = p^2K^2
\end{equation}
i.e.\ $K^2$ is related to the moment of inertia $B$. Now the
irreducibility demanded on the representation level implies that
${\mit\Lambda}$ and $K$ are multiplies of the identity. To obtain the
standard quantum-mechanical limit  we can choose
\begin{equation}
{\mit\Lambda} = I, \quad K = {1 \over {\sqrt{2B}}}I.
\end{equation}
The Hamiltonian form of the Heisenberg equations of motion
reads:
\begin{eqnarray}
&&\dot{{\mit\Lambda}} = \dot{K} = 0 \nonumber \\
&&\dot{x} =
-{i \over {2B}}x(2p-\hbar) \label{qm.heisenberg}\\
&&\dot{p} = 0 \nonumber
\end{eqnarray}
Eqs.\ (\ref{qm.heisenberg}) have the well known solution
\begin{equation}
p(t) = p_0, \quad x(t) = x_0e^{-{i \over {2B}}(2p_0 - \hbar)t}
\label{qm.heisenberg.sol}
\end{equation}
where $p_0$ denotes initial angular momentum and $x_0$ denotes
the initial position of a particle. Equations
(\ref{qm.heisenberg}) as well as (\ref{qm.heisenberg.sol}) are
just  identical to the equations obtained from
considerations of the algebra  $\cal H$, therefore  the
algebras ${\cal H}'$  and $\cal  H$ describe the same physical
situation. In the next section however we will see that the
algebra $\H'$ will be suitable to the non-commuative extension
of the described quantum mechanical model.

\section{$q$-deformed quantum particle on a circle}

Quantum cylinder is defined as a free involutive algebra
generated by the identity $I$, unitary $x$ and hermitean $p$
modulo the relations
\begin{equation}
xp = qpx, \quad x^* = x^{-1}, \quad p^* = p,
\label{qcyll}
\end{equation}
where $q$  is a positive real number. This space can be
considered as a phase-space of the $q$-deformed particle on a
circle. To  quantise this system we have to replace the first of
equations (\ref{qcyll}) by the equation
\begin{equation}
xp =qpx + \hbar {\mit\Lambda} x.
\end{equation}
It means that we have to deform consistently the extended  algebra
of observables $\H'$ to the algebra $\H_{q\varepsilon\xi}$ i.e.\ we have to
deform the ideal $J(I,x,p,K,{\mit\Lambda})$ to the ideal
$J_{q\varepsilon\xi}(I,x,p,K,{\mit\Lambda})$ in such a way that  both $K$ and
${\mit\Lambda}$  are  no longer commutative in the algebra
$\H_{q\varepsilon\xi}
 =
{\C}[I,x,p,K,{\mit\Lambda}]/J_{q\varepsilon\xi}$. This can be done by the
replacement of equations (\ref{def.j}) by the set of the
following relations
\begin{eqnarray}
xp & = & qpx + \hbar {\mit\Lambda} x \nonumber \\
x {\mit\Lambda} & = & \varepsilon {\mit\Lambda} x \nonumber \\
p {\mit\Lambda} &  = & {\mit\Lambda} p \nonumber \\
xK &  = & \xi Kx \label{def.qj} \\
pK & = & Kp \nonumber \\
K{\mit\Lambda} & = & {\mit\Lambda} K, \nonumber
\end{eqnarray}
where all parameters $q, \varepsilon , \xi$ are real and positive.
We use the same $^*$ structure as in the commutative case,
i.e.\ $p,K,{\mit\Lambda}$ are hermitean and $x$ is unitary. Now we can
consider the unitary time evolution of the system with the
Hamiltonian:
\begin{equation}
H = p^2K^2 + V(K, {\mit\Lambda})
\label{q.hamiltonian}
\end{equation}
where $V$ is an arbitrary element of ${\cal H}_{q\varepsilon\xi}$
constructed only from $K$ and ${\mit\Lambda}$. It leads to the Hamilton
equations of the form
\begin{eqnarray}
\dot{{\mit\Lambda}}&=&\dot{K}=0
\nonumber \\
\dot{x} & = & ix[-2\varepsilon ^{-1} {\mit\Lambda} K^2p + \hbar
\varepsilon ^{-2} ({\mit\Lambda} K)^2 + \hbar^{-1}
(V(qK,\varepsilon^{-1}{\mit\Lambda} ) - V(K,{\mit\Lambda}))]
\label{qheisenberg}\\
\dot{p} &= &0. \nonumber
\end{eqnarray}
Here we have used  the natural condition, that $\dot{x}$ is linear in the
moment
um $p$. This condition is satisfied if
\begin{equation}
\xi = q^{-1}. \label{ksi}
\end{equation}
This reduces algebra $\H_{q\varepsilon\xi}$ to the algebra
$\H_{q\varepsilon}$ given by the following relations:
\begin{eqnarray}
xp &=& qpx + \hbar {\mit\Lambda} x \nonumber \\
x{\mit\Lambda} &=& \varepsilon {\mit\Lambda} x \nonumber \\
p{\mit\Lambda} &=& {\mit\Lambda} p \nonumber \\
xK &=& q^{-1} Kx \label{def.qj1}\\
pK &=& Kp \nonumber \\
K{\mit\Lambda} &=& {\mit\Lambda} K. \nonumber
\end{eqnarray}
Note that the function $V(K,{\mit\Lambda})$ which plays a role of
a scale of energy in the non-deformed case now comes into
equations of motion, giving a correction to the velocity of
a particle.  In principle this correction can be quite large
since it is proportional to the inverse of the Planck constant.
If we demand the existence of classical limit ($\hbar =0$) we
have to estimate $q = 1 + c_1\hbar + O(\hbar^2)$, $\varepsilon =
1  + c_2\hbar +O(\hbar^2)$. One can interpret this term as an
internal or not related to the motion contribution to the
angular velocity.  We will see in Section 5  that the
dependence of $\dot{x}$ on $V(K,{\mit\Lambda})$ vanishes, when we assume that
the Hamiltonian is invariant under the time inversion. Equations
(\ref{qheisenberg}) have the solution
\begin{equation}
p(t)=p_0, \quad x(t)=x_0e^{i[-2\varepsilon ^{-1} {\mit\Lambda} K^2p_0 +
\hbar
\varepsilon ^{-2} ({\mit\Lambda} K)^2 + \hbar^{-1}
(V(qK,\varepsilon^{-1}{\mit\Lambda} ) - V(K,{\mit\Lambda}))]t}
\end{equation}
from what we see immediately, that the time evolution of the
system is given by the unitary operator $U=\exp
({i\over\hbar}Ht)$.

\section{Representations of $\H_{q\varepsilon}$ in the Hilbert space}

In this section we will find the Hilbert space of the
representations of the $q$-deformed algebra of observables $\H_{q\varepsilon}$.
 The simplest way
to do this is to define the action of the
operators $x,p, {\mit\Lambda}$ and $K$ on the orthonormal set of
vectors. The operation $^*$ is represented by the hermitean
conjugation. Since $K,p$, and ${\mit\Lambda}$ are hermitean and
commute, they can be diagonalised simultanously. Let us denote
the eigenvectors of $p,K$ and ${\mit\Lambda}$ by $| k,\kappa,
\lambda\rangle$ where $k,\kappa,\lambda$ belong to the spectrum of
$p,K,{\mit\Lambda}$ respectively, i.e.\
\begin{eqnarray}
p | k,\kappa,\lambda \rangle & = & k | k,\kappa,\lambda \rangle \nonumber \\
K | k,\kappa,\lambda \rangle & = & \kappa | k,\kappa,\lambda \rangle \\
{\mit\Lambda} | k,\kappa,\lambda \rangle & = & \lambda | k,\kappa,\lambda
\rangle.
\nonumber
\end{eqnarray}
By means of equations (\ref{def.qj1}) we see that
\begin{eqnarray}
x^n | k,\kappa,\lambda\rangle & = & | q^{-n}(k-n\hbar\varepsilon^{-1}\lambda),
q^n \kappa,\varepsilon^{-n}\lambda \rangle \label{rep.x}\\
& = & | k',\kappa ',\lambda ' \rangle \nonumber
\end{eqnarray}
are again eigenvectors of $p,K$ and ${\mit\Lambda}$. Therefore
the Hilbert space of representations of the described
system is spanned by a lattice of vectors defined by (\ref{rep.x}). Let
us now classify representations of $\H_{q\varepsilon}$. We normalize $K$
in such a way that it becomes $I$ in a non-deformed case. Assume first that
$q \geq 1$ and $\varepsilon \geq 1$. Then each set of numbers
$\kappa_0, \lambda_0, k_0$ such that
\begin{equation} q^{-\h } \leq
\kappa_0 < q^\h , \quad \varepsilon^{-\h } \leq \lambda_0 < \varepsilon^\h ,
\quad 0 \leq k_0 < \hbar \varepsilon^{-1}\lambda_0
\end{equation}
defines the irreducible representation of $\H_{q\varepsilon}$.  Similarly if
we assume that $q <1$ and $\varepsilon \geq 1$, we obtain that the
irreducible representations of $\H_{q\varepsilon}$ can be labelled by the
numbers $k_0,\kappa_0 ,\lambda_0$ such that
\begin{equation}
q^\h \leq \kappa_0 < q^{-\h} , \quad \varepsilon^{-\h } \leq \lambda_0
< \varepsilon^\h ,
\quad - \hbar \varepsilon^{-1}\lambda_0 < k_0 \leq 0
\end{equation}
The same classification  can be repeated for $\varepsilon <1$ but
now $\varepsilon^\h \leq \lambda_0 <\varepsilon^{-\h}$.

Immediately from the equation (\ref{rep.x}) we see that the angular
momentum $p$ is quantised. Moreover the classification of
irreducible representations suggests that in general free particle on a
$q$-deformed circle has anyonic rather than usual Bose-Fermi
statistics.

Let us now put $V(K,{\mit\Lambda}) = 0$. Using equation (\ref{rep.x}) we
can obtain the energy spectrum of a free particle moving along
a quantum circle
\begin{equation}
E_n = (k_0-n\hbar \varepsilon^{-1}\lambda_0)^2\kappa_0^2.
\end{equation}

\section{Symmetries}
It is an easy exercise to check that the
algebra $\H_{q\varepsilon}$ is invariant under the transformation
$x\to -x$, $p
\to p$, $K\to K$, ${\mit\Lambda}\to {\mit\Lambda}$.
This transformation corresponds to the space inversion in the
non-deformed case. The representation (\ref{rep.x}) is obviously
invariant under this transformation. In the non-deformed case
however the algebra (2)  is also invariant under the time inversion $x
\to x^{-1}$, $p\to -p$ which is no longer a
symmetry of $\H_{q\varepsilon}$. We can try to find the symmetry of
$\H_{q\varepsilon}$ which generalizes the time-inversion. To do
this let us consider the general, anti-unitary transformation
$\T$, $\T^2=I$ defined as
\begin{eqnarray}
\T x\T^{-1} = x^{-1}, \quad \T p\T^{-1} = -pf(K,{\mit\Lambda}),\\
\T K\T^{-1} = g(K,{\mit\Lambda}), \quad \T {\mit\Lambda} \T^{-1}=
h(K,{\mit\Lambda})\nonumber
\end{eqnarray}
where $f,g,h$ are arbitrary elements of $\H_{q\varepsilon}$ depending only on
$K,{\mit\Lambda}$. The algebra $\H_{q\varepsilon}$ is invariant under this
transformation if and only if $\varepsilon = q$ and $\T$ has the
form
\begin{eqnarray}
\T x\T^{-1} & = & x^{-1} \nonumber \\
\T p\T^{-1} & = & -pf(K,{\mit\Lambda}) \nonumber \\
\T K\T^{-1} & = & f^{-1}(K,{\mit\Lambda}) K \\
\T {\mit\Lambda} \T^{-1} & = & f(K,{\mit\Lambda}) {\mit\Lambda} \nonumber
\end{eqnarray}
where function $f$ fulfils the condition
\begin{equation}
q^2f(q^{-1}K, q{\mit\Lambda}) = f(K,{\mit\Lambda}).
\label{eqn.f}
\end{equation}
The algebra $\H_{q\varepsilon}$ reduces now to $\H_q$ with only one
deformation parameter $q$ and the commutation relations
\begin{eqnarray}
xp &=& qpx + \hbar {\mit\Lambda} x \nonumber \\ x{\mit\Lambda} &=& q
{\mit\Lambda}
x \nonumber \\ p{\mit\Lambda} &=& {\mit\Lambda} p \nonumber \\ xK &=& q^{-1}
Kx \label{def.qj2}\\ pK &=& Kp \nonumber \\ K{\mit\Lambda} &=& {\mit\Lambda}
K. \nonumber
\end{eqnarray}
{}From the quations (\ref{def.qj2})  it follows  that $I$  and
$K{\mit\Lambda}$ generate the center of $\H_q$. Focusing on the
irreducible representations of $\H_q$ we can put
\begin{equation}
K = c{\mit\Lambda}^{-1},
\end{equation}
where the constant $c$ can be derived from the classical limit,
i.e.\ $c = (2B)^{-\h}$. Using this identification and the
classical limit we can easily solve equation (\ref{eqn.f}),
namely
\begin{equation}
f(K) = {\mit\Lambda}^{-2},
\end{equation}
i.e.\
\begin{equation}
\T x\T^{-1} = x^{-1} , \quad \T  p \T^{-1} = -p{\mit\Lambda}^{-2},
\quad \T{\mit\Lambda}\T^{-1} = {\mit\Lambda}^{-1}.
\end{equation}
In this case the action of the position operator $x$ on the
basis $ | k,\lambda \rangle$ takes the form
\begin{equation}
x^n | k, \lambda \rangle = | q^{-n} (k -  n\hbar q^{-1}\lambda ),
q^{-n}\lambda \rangle.
\end{equation}
An irreducible representation of $\H_q$ contains vectors numbered
by the knots of the lattice generated by $x^n$,
\[ \{ q^{-n}(k-  n\hbar q^{-1}\lambda), q^{-n} \lambda \} \]
 with integer $n$. Now
\[ \T | k,\lambda\rangle = | -\lambda^{-2}k,\lambda^{-1} \rangle  \]
thus the lattice should contain point $(q^{-n}(k-n\hbar
q^{-1}\lambda),q^{-n}k)$ together with the point
$(-q^{-n}\lambda^{-2}(k-n\hbar q^{-1}),q^n\lambda^{-1})$. It
means that we have two kinds of irreducible representation. One
is given by the choice $\lambda_0 =1$, $k_0=0$ while the second one
can be generated from $\lambda_0= q^\h$, $k_0 =\h q^\h \hbar$. They
correspond to Bose and Fermi statistics respectively like in the
non-deformed case.

If we now consider the motion given by the Hamiltonian
(\ref{q.hamiltonian}) and demand that $\T H \T^{-1} = H$ we will
see that $V({\mit\Lambda} ) = {\rm const}$. $I$ and the equations of motion
will take the form
\begin{eqnarray}
\dot{{\mit\Lambda}} & = & 0 \nonumber \\
\dot{x} & = & ix [-2q^{-1}{\mit\Lambda} K^2p + \hbar q^{-2}({\mit\Lambda} K)^2]
\\
\dot{p} & = & 0. \nonumber
\end{eqnarray}
i.e.\ the additional term $V$ does not contribute to the angular
velocity of the particle.

If we now introduce new variables:
\begin{equation}
P = qp{\mit\Lambda}^{-1}, \quad X = x
\label{trans.class}
\end{equation}
then we obtain the canonical commutation relation $[X,P] =
\hbar X$ of the standard quantum mechanics on a circle. This is in a
contrast to the case $\varepsilon \neq q$, where such a
reparametrization is not possible. Note however that even in the
case $q = \varepsilon$ the deformed theory can not be treated just
as a standard quantum mechanics with non-commutative moment of
inertia, since transformation (\ref{trans.class}) is not
unitary.

\section{Conclusions}
We described the $q$-deformation of the quantum mechanics of a
particle on a circle. To describe the unitary time evolution of
the system we had to deform the algebra of observables leaving
the Heisenberg equations of motion unchanged. We were able to
reduce number of free parameters requiring the existence of
classical limit and finally the symmetry  of irreducible
representations on the time-inversion. In this last case we
found a non-unitary transformation of variables which allowed us
to replace deformed cannonical commutation relations by the
standard ones. This possibility was the straightforward
consequence of the additional symmetry (time-inversion) in a
contrast to the situation described in \cite{rembielinski1}.

\vspace{14pt}

\section*{Acknowledgement} T. Brzezi\'nski would like to thank St.\
John's College, Cambridge for a Benefactors' Studentship.


\baselineskip 14pt

\begin{thebibliography}{10}

\bibitem{alekseev1}
A.Yu. Alekseev, L.D. Faddeev:
{\em Commun.\ Math.\ Phys.} {\bf 141}:413, (1991);
L.D. Faddeev:
Quantum symmetry in conformal field theory by Hamiltonian methods,
{\em Helsinki University preprint}, (1991).

\bibitem{arefeva1}
I.Ya.\ Aref'eva, I.V. Volovich:
Quantum group particles and non-Archimedean geometry,
{\em Phys.\ Lett.}, {\bf B268}:179, (1991).

\bibitem{connes2}
A. Connes:
{\em G\'eometrie non commutative},
Inter Editions, (1990).

\bibitem{dimakis1}
A. Dimakis, F. M\"uller-Hoissen:
Quantum mechanics as a non-com\-mu\-ta\-tive symplectic geometry,
{\em G\"ottingen Univeristy preprint}, (1991).

\bibitem{manin2}
Yu.I. Manin:
{\em Quantum groups and non-commutative geometry},
Montreal Notes, (1989).

\bibitem{rembielinski1}
J. Rembieli\'nski:
Non-commutative dynamics,
{\em \L\'od\'z Univeristy preprint}, (1992).

\bibitem{rembielinski2}
J. Rembieli\'nski:
Non-commutative relativistic kinematics,
{\em Phys.\ Lett.} {\bf B287}:145, (1992).

\bibitem{wess2}
J. Wess:
A talk given at XIX ICGTMP Salamanca (Spain), 1992 and
J. Schwenk, J.~Wess:
A $q$-deformed quantum mechanical toy model,
{\em Phys. Lett.} {\bf B291}:273, (1992).

\end{thebibliography}
\end{document}